\documentclass[prl,twocolumn,superscriptaddress,showpacs,psfig,showkeys,nofootinbib]{revtex4}
\usepackage{amsmath}
\usepackage{amssymb}
\usepackage{amsfonts}
\usepackage{graphicx}
\usepackage{dcolumn}
\usepackage{hyperref}
\usepackage{epstopdf}
\usepackage{color}
\usepackage[utf8]{inputenc}
\usepackage{amsthm}
\usepackage{fontenc}
\usepackage{bm}
\RequirePackage{slashed}\usepackage{cancel}
\usepackage[normalem]{ulem}

\usepackage{array,booktabs}
\newcolumntype{C}{>{$\displaystyle}c<{$}}

\usepackage{soul}

\newcommand \Pomeron {I\!\!P}

\begin{document}


\title{Coherent $J/ \psi$ electroproduction on $^4$He and $^3$He at the Electron-Ion Collider: 
\\
probing nuclear shadowing one nucleon at a time}

\author{Vadim Guzey}
\affiliation{National Research Center ``Kurchatov Institute", Petersburg Nuclear Physics Institute (PNPI),Gatchina, 188300, Russia}
\author{Matteo Rinaldi} 
\affiliation{ Dipartimento di Fisica e Geologia,
Universit\`a degli Studi di Perugia and Istituto Nazionale di Fisica Nucleare,
Sezione di Perugia, via A. Pascoli, I - 06123 Perugia, Italy}
\author{Sergio Scopetta}
\affiliation{ Dipartimento di Fisica e Geologia,
Universit\`a degli Studi di Perugia and Istituto Nazionale di Fisica Nucleare,
Sezione di Perugia, via A. Pascoli, I - 06123 Perugia, Italy}
\author{Mark Strikman}
\affiliation{Pennsylvania  State  University,  University  Park,  PA,  16802,  USA} 
\date{\today}
\author{Michele Viviani}
\affiliation{
INFN-Pisa, 56127 Pisa, Italy} 
\date{\today}

\begin{abstract}

While the phenomenon of gluon nuclear shadowing at small $x$ has been getting confirmation in QCD analyses of various LHC 
measurements involving heavy nuclei, it has not been possible so far to establish experimentally the number of target nucleons responsible for nuclear
shadowing in a given process. To address this issue, we study coherent $J / \psi$ electroproduction
on $^4$He and $^3$He in the kinematics of a future electron-ion collider
and show that this process has the power to disentangle the contributions of the interaction with
a specific number of nucleons $k$, in particular,
with two nucleons at the momentum transfer $t \neq 0$. 
We predict a dramatic shift of the $t$-dependence of the differential cross section toward smaller values of $|t|$ due to a non-trivial correlation between 
$x$ and $k$.
This calculation, which makes use for the first time of realistic wave functions, provides a stringent test of models of nuclear shadowing and a novel probe of the 3D imaging of gluons in light nuclei.
In addition, thanks to this analysis, unique information on the real part of the corresponding scattering amplitude could be accessed.

\end{abstract}

\maketitle

\textbf{\textit{Introduction}}.
Studies of nuclear shadowing have a long history~\cite{Frankfurt:1988nt,Arneodo:1992wf,Piller:1999wx,Armesto:2006ph,Frankfurt:2011cs}.
In quantum mechanics and in the eikonal limit, it is manifested in the total hadron-nucleus cross section being smaller 
than the sum of 
individual hadron-nucleon cross sections.
In essence, this is due to 
simultaneous interactions of the projectile with $k \geq 2$
nucleons of the nuclear target, leading to a reduction (shadowing) of the total cross section. 
In this framework the interaction of the projectile with a nucleus is described by a sum of diagrams corresponding to
the potential interaction with individual nucleons, giving rise to the Glauber model~\cite{Glauber:1955qq,Glauber:1970jm}.
However, it was demonstrated by Mandelstam~\cite{Mandelstam:1963cw} and Gribov~\cite{Gribov:1968jf} that the contribution of eikonal diagrams in quantum field theory models tends to zero at high energies because, qualitatively, there is not enough time between interactions with two nucleons for the projectile to transform back into itself.
As a result, shadowing in the high energy limit is determined by the totality of diffractive interactions of the projectile in different configurations~\cite{Gribov:1968jf}.

More recently the issue of nuclear shadowing in hard processes with nuclei was discussed in the context of
modifications of nuclear parton distribution functions (PDFs) and an eventual onset of the regime of high gluon densities (saturation), which are relevant for the physics programs of the Large Hadron Collider (LHC)~\cite{Salgado:2011wc,Citron:2018lsq} 
and of a future Electron-Ion Collider (EIC)~\cite{Accardi:2012qut}.
Specifically, combining the Gribov-Glauber approach to nuclear shadowing with the 
collinear QCD factorization theorems for diffractive and inclusive leading-twist processes in deep inelastic scattering (DIS), a large leading-twist (LT) gluon nuclear shadowing at small $x$ was predicted~\cite{Frankfurt:1998ym} ($x$ is the nucleus momentum fraction carried by the gluons).
It was later confirmed by analyses of coherent photoproduction of charmonia in ultraperipheral collisions
(UPCs) of heavy ions at the LHC, which showed~\cite{Guzey:2013xba,Guzey:2013qza,Guzey:2020ntc} that 
$R_g(x=6 \times 10^{-4}-10^{-3},\mu^2 \approx 3\ {\rm GeV}^2)=g_A(x,\mu^2)/[A g_N(x,\mu^2)] \approx 0.6$ for lead nuclei
($g_A(x,\mu^2)$ and $g_N(x,\mu^2)$ refer to the gluon density in the nucleus and the nucleon, respectively).
Alternative calculations performed in the eikonal dipole models, where nuclear shadowing is a higher twist effect, 
lead to a somewhat weaker shadowing; see, e.g., Ref.~\cite{Lappi:2013am}.

Exclusive electroproduction of $J/\psi$ probes directly the gluon density of the target~\cite{Ryskin:1992ui,Brodsky:1994kf}.
The large magnitude of nuclear shadowing means that 
gluons in heavy nuclei probed in this process likely belong to more than one nucleon of the target
due to their overlap in the transverse plane~\cite{Frankfurt:2011cs,Guzey:2016qwo}.
However, it is difficult to discriminate between different mechanisms of shadowing using scattering off heavy nuclei since the correction in this case is a result of summing a sign-alternating and slowly converging series.
In addition, measurements of coherent scattering with the momentum transfer squared $t \ne 0$ are challenging 
because of a steep $t$ dependence.
As a consequence, it is impossible to establish the exact number of target nucleons involved in the process, which hinders an access to 
information on the gluon dynamics in nuclei.

In this Letter, starting from the successful description of the effect of nuclear shadowing for heavy nuclei,
we propose an alternative, complementary strategy of studying coherent production of $J/\psi$ in DIS off
$^4$He and $^3$He light nuclei
in the $Q^2 \to 0$ (quasireal photon) limit, which should be feasible at the 
EIC. 
In this case, $x=M_{J/\psi}^2/W^2$, where $M_{J/\psi}$ is the vector meson mass and $W$ is the invariant photon-nucleon energy.

Using specific features of their response functions, namely, the presence of a zero in the one-body form factor (ff) 
at moderate 
$\sqrt{|t|} =0.7$ GeV/c for $^4$He and $\sqrt{|t|} =0.8$ GeV/c for $^3$He,
we argue that it is possible to separate the contributions to nuclear shadowing coming from the interaction with two and three nucleons of the nuclear target.
Besides, the ions under investigation have
no excited states, so that it is easy to select coherent events.
This is based on an old idea proposed initially in Ref.~\cite{Levin:1975pc}.
Indeed, since
 the differential $p+ ^{4}{\rm He} \to p+^{4}{\rm He}$
cross section does not present a minimum at $-t \simeq 0.6$ GeV$^2$, where 
the $^4$He charge ff has a minimum,
it has to be dominated by effects beyond the impulse approximation (IA), namely, by the interaction with several nucleons leading 
to nuclear shadowing. Supplementing this with accurate calculations of one-, two-, and three-body ffs based on 
exact solutions of the Schr\"odinger equation with realistic potentials, which have greatly improved in the last decade,
we make predictions for the $d\sigma_{\gamma^{\ast}+^{4}{\rm He} \to J/\psi+^{4}{\rm He}}/dt$
and $d\sigma_{\gamma^{\ast}+^{3}{\rm He} \to J/\psi+^{3}{\rm He}}/dt$ differential cross sections in a broad range of $t$, 
including the region of finite $t$, where IA is largely suppressed and  
the cross sections are unambiguously sensitive to the contributions of the interaction with exactly two nucleons of the target (the contribution of interactions with three and four nucleons near the minimum and for lower $|t|$
is numerically small).

Note that runs with polarized $^3$He beams are planned at EIC in order to study the neutron spin structure~\cite{Accardi:2012qut}.
 
In the case of electron-deuteron scattering, the IA induced by the quadrupole ff dominates up to large $|t|$. 
Hence, we do not discuss this reaction here. However, in a long run, if polarized deuteron beams for an EIC 
become available, experiments using such beams would provide an independent measurement of the double scattering amplitude
(the interaction with $k=2$ nucleons).
At the same time, the strategy discussed here is probably the only one allowing one to measure nuclear 
shadowing in light nuclei at colliders, since its effect for the total electron-nucleus cross sections is a few percent at most~\cite{Frankfurt:2003jf}.
Its smallness can be readily seen by examining our predictions in Figs.~\ref{f_tre}--\ref{f_5},
where the difference between the IA and full results at $t=0$ is twice the shadowing effect for the total
cross section.

\begin{figure}
\includegraphics[scale=0.55,angle=0]{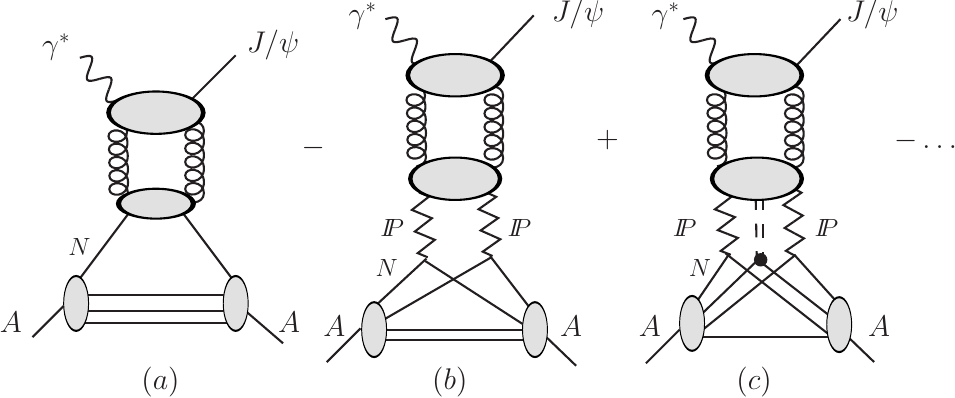} 
\caption{Multiple scattering series for
 $\gamma^{\ast}+A \to J/\psi+A$ scattering amplitude. Panels (a), (b), and (c) correspond to
the interaction with $k=1$, 2, and 3 target nucleons, respectively.
The zigzag lines labeled $\Pomeron$ denote diffractive exchanges; the solid
blob in panel (c) stands for the interaction with cross section $\sigma_3(x)$.}
\label{fig:GG}
\end{figure}

\textbf{\textit{Multiple scattering formalism for coherent electroproduction of $J/\psi$ on light nuclei}}.
As already explained, at high energies projectiles interact coherently with all nucleons of the nuclear target. The contributions to the nuclear scattering amplitude corresponding to the interaction with $k=1, 2, 3, \dots$
nucleons of the target is shown in Fig.~\ref{fig:GG};    they
interfere destructively
leading to the suppression of the nuclear cross section (nuclear shadowing)~\cite{Glauber:1955qq}.  
In the Gribov-Glauber approach to nuclear shadowing~\cite{Gribov:1968jf}, the contribution of the interaction with $k=2$ nucleons
is unambiguously given by the diffractive (elastic) cross section on the nucleon. 
 At the same time, the contributions corresponding to $k \geq 3$
 nucleons cannot be in general expressed in terms of diffraction on the nucleon and, hence, need to be modeled.
 A convenient way to do it
 is offered by the Good-Walker formalism of eigenstates of the 
 scattering operator~\cite{Good:1960ba,Blaettel:1993ah}, which allows one to characterize the interaction with $k$
 nucleons by the $k$th moment $\langle \sigma^k \rangle=\int d\sigma P_h(\sigma) \sigma^k$.
Here, the distribution $P_h(\sigma)$ parametrizes the hadronic structure of the virtual photon and gives the probability for the photon to interact with a nucleon with the cross section $\sigma$.
 
\begin{figure}
\begin{center}
\includegraphics[scale=0.3,angle=0]{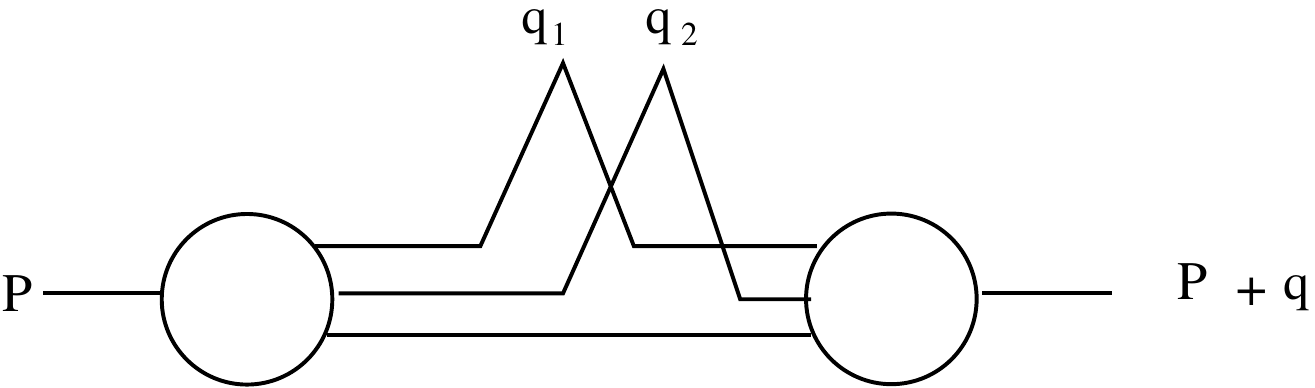} 
\end{center}
\caption{A sketch of the two-body ff $\Phi_2 $; see Eq.~(\ref{cinque}). }
\label{f_due}
\end{figure}

\begin{figure}[t]
\begin{center}
\hspace*{-0.7cm}\includegraphics[scale=0.4]{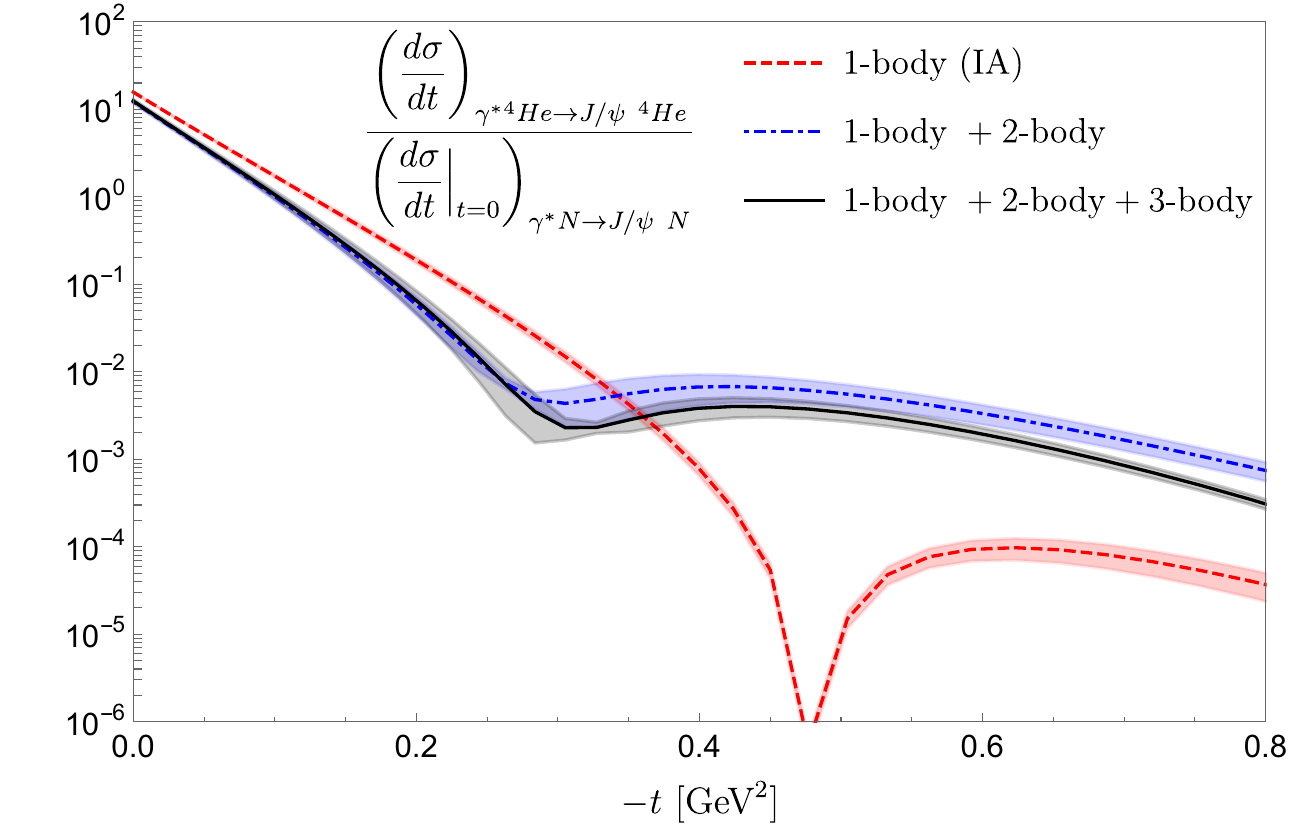}
\end{center}
\caption{
Ratio of the differential
cross section for $J/\psi$ coherent production on
$^4$He to the same quantity for the nucleon target at $t=0$
as a function of $-t$
at $x=10^{-3}$.
Relative errors of 10\% and 15\% have been considered on
the quantities $B_0$ and $\sigma_2$, respectively (see text and the Supplemental Material).}
\label{f_tre}
\end{figure}

\begin{figure}[t]
\begin{center}
\hspace*{-0.7cm}\includegraphics[scale=0.44]{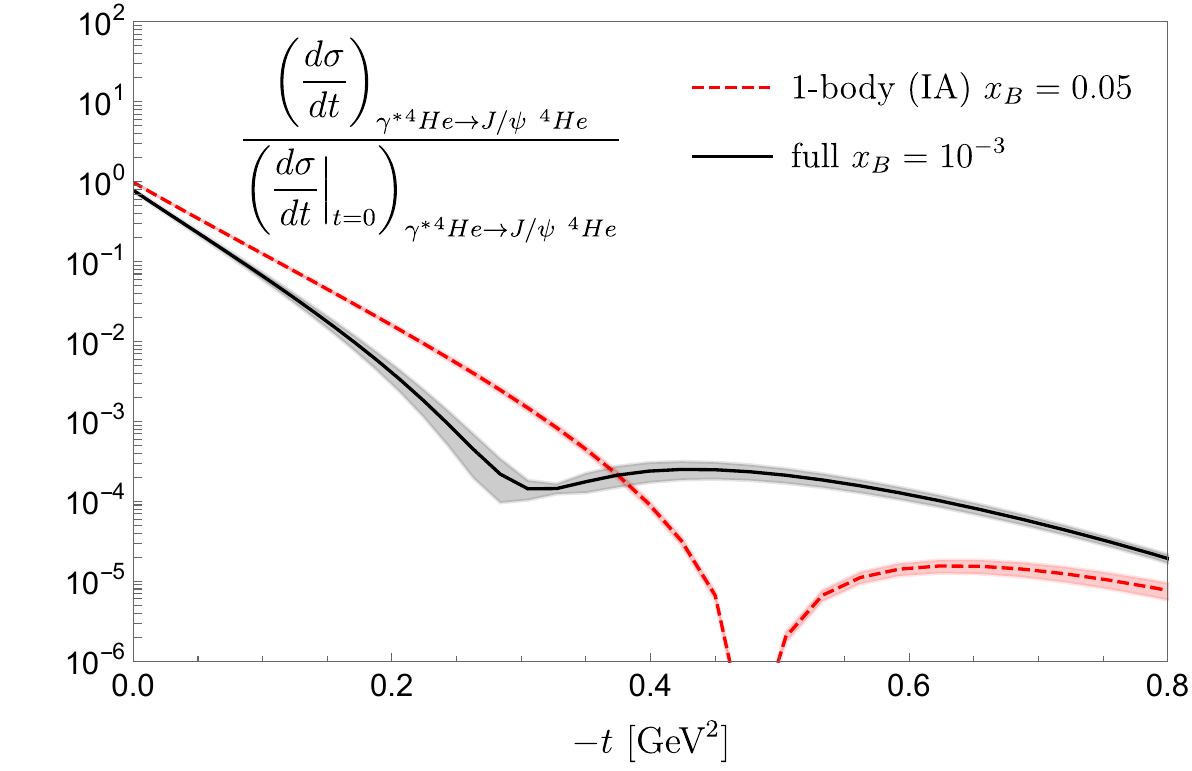}
\end{center}
\caption{
Ratio of the differential
cross section for $J/\psi$ coherent production on
$^4$He to the same quantity at $t=0$
as a function of $-t$: the IA result at $x=0.05$
is compared to the full one at $x=10^{-3}$.
Relative errors of 10\% and 15\% have been considered on
the quantities $B_0$ and $\sigma_2$, respectively (see text and
the Supplemental Material).}
\label{fig_4}
\end{figure}

\begin{figure}[t]
\begin{center}
\hspace*{-0.7cm} \includegraphics[scale=0.4]{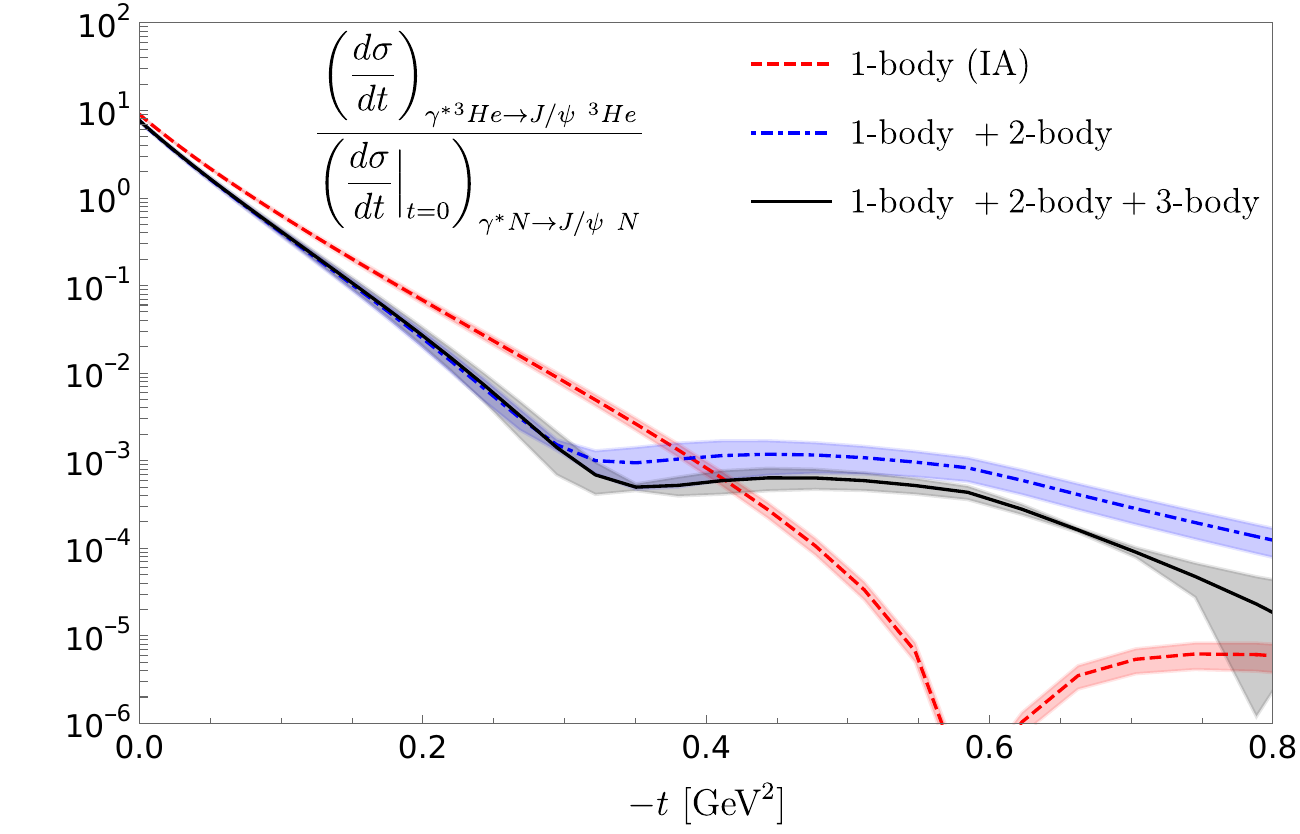}
\end{center}
\caption{
Ratio of the differential
cross section for $J/\psi$ coherent production on
$^3$He to the same quantity for the nucleon target at $t=0$
as a function of $-t$ at $x=10^{-3}$.
Relative errors of 10\% and 15\% have been considered on
the quantities $B_0$ and $\sigma_2$, respectively (see text and
the Supplemental Material).}
\label{f_5}
\end{figure}

In this approach, 
while the $\gamma^{\ast}A \to J/\psi A$ scattering amplitude is expressed in terms of the gluon generalized parton distribution (GPD),  the ratio of the nuclear and nucleon cross sections very weakly depends on the skewness.
Thus, 
the differential cross section of electroproduction (photoproduction) of $J/\psi$ mesons on a nucleus $A$ can be 
written in terms of the $d \sigma_{\gamma^{\ast} N \rightarrow J/\psi N}/dt$ cross section on the proton at $t=0$
in the following form:
\begin{eqnarray}
\frac{d \sigma_{\gamma^{\ast} A \rightarrow J/\psi A }}{dt}(t)
& = &
\frac{d \sigma_{\gamma^{\ast} N \rightarrow J/\psi N }}{d t}(t=0)
\nonumber
\\
& \times &
\left |F_1(t)e^{[B_{0}(x)/2]t}+
\sum_{k=2}^A F_k(t) \right |^2 \,,
\label{uno}
\end{eqnarray}
where the first term is proportional to the IA,
$F_1(t) = A {{\Phi_1(q)}}$,
and the second term gives the contribution of the interaction with $2 \leq k \leq A$ nucleons~\cite{Levin:1975pc}, 
\begin{eqnarray}
    F_{k}(t) & = & \left (-
    \frac{1}{8 \pi^2} \right)^{k-1}
    \binom{A}{k} \frac{\langle  \sigma^k \rangle }{\langle \sigma \rangle } 
\frac{(1 -i \eta)^k}{1- i \eta_0}\
    \int
    \prod_{l=1}^k d^2 \vec{q}_l 
\nonumber
\\
& \times & \, {{f(q_l)}} \,
{{\Phi_k(\vec{q},\vec{q}_l)}} \,
\delta^2 \left ( \sum_{l=1}^k \vec{q}_l -\vec{q} \right)
\,,
\label{tre}
\end{eqnarray}
where all the transferred momenta $\vec{q}_i$ have been taken purely transverse;
$t=-|\vec{q}|^2$.
The nuclear structure is taken into account via the $k$-body ffs $\Phi_k$
\begin{eqnarray}
{{\Phi_k(\vec q_{1}, ... \vec q_k)}}
& = & 
\int 
\prod_{i=1}^A  
\left \{ \frac{d \vec p_i}{(2 \pi)^3} \right \} 
\nonumber\\
 & \times  & 
\psi^{\ast}_{P}(\vec p_1+ \vec q_{1 }, ... \vec p_k+ \vec q_{k },..., \vec p_A) \, 
\nonumber
\\
& \times &
\psi_P(\vec p_1, 
..., \vec p_k,... \vec p_A) \, \delta\left (\sum_{i=1}^A \vec p_i \right ) \,,
\label{cinque}
\end{eqnarray}
which represent the probability amplitude for $k$ nucleons in the nucleus $A$ to 
interact with the probe absorbing the momenta $\vec{q}_1,\cdots,\vec{q}_k$ and then going back
to the nucleus, which recoils elastically with a total momentum
transfer $\vec{q}=\sum_k \vec{q}_k$. Furthermore, $\psi_P$ is the intrinsic 
nuclear wave function 
with total momentum $P$.
A pictorial representation of the two-body ff $\Phi_2$ is shown in Fig.~\ref{f_due}.
One should notice that Eq.~(\ref{cinque}) represents an integral over $3^{A-1}$ variables
of complicated wave functions for each ${q_i,...q_k}$ set of values.
Thus, to evaluate accurately Eq.~(\ref{tre}), the function $\Phi_k$ is needed 
on an enough dense grid of these $k$ variables. As a consequence,
the computation time grows dramatically with $A$ and $k$.

In Eqs.~(\ref{uno}) and (\ref{tre}), the $t$ dependence of the $\gamma^{\ast} N \rightarrow J/\psi N$ and soft hadronic 
fluctuation-nucleon amplitudes is characterized by the slope $B_0(x)$ and the factor $f(q_l)$, respectively;
$\eta \neq \eta_0$ are the ratios of the real to imaginary parts of these amplitudes.

Note that the multiple scattering formalism presented in this section provides 
a good description of the data on proton-$^4$He elastic scattering~\cite{Bujak:1978rla}.

\textbf{\textit{Results for $^4$He and $^3$He}}.
Below, we specify input for our calculations and show our predictions for the differential cross sections
of $J/\psi$ electroproduction on $^4$He and $^3$He in generic kinematics of an EIC. 
We start with $^4$He, for which preliminary results have been presented
in the EIC Yellow Report~\cite{AbdulKhalek:2021gbh}.

The $k$-body ffs $\Phi_1$, $\Phi_2$, and $\Phi_3$ have been calculated using a realistic nuclear wave function 
evaluated along the lines of Ref.~\cite{Marcucci:2019hml} and using the N$^4$LO chiral potential
of Ref.~\cite{Entem:2017gor}
with a cutoff of 500 MeV and three-body forces. Within this approach, the binding energy of 
$^4$He turns out to be $B(^4{\rm He})=-28.15$ MeV, which is very close to the experimental value.
A consistency check of the numerics is provided by the  relation
$\Phi_2(\vec q_2=0,\vec q_{1}) =
\Phi_1(\vec q_{1})$,  
which has been successfully tested.

A cumbersome realistic calculation of the $\Phi_4$ ff has not been performed since
a very small contribution is predicted. This expectation is supported by an estimate carried out within a 
harmonic oscillator shell model and
treating properly 
the center-of-mass motion,
as suggested in Ref.~\cite{Levin:1975pc}.
We verified that this procedure 
reproduces the relative sizes
of the $\Phi_2$ and $\Phi_3$ contributions obtained within the realistic analysis 
reasonably well.

In the leading logarithmic approximation of perturbative QCD, coherent electro- and photoproduction of $J/\psi$ on nuclei probes
the gluon density of the target~\cite{Guzey:2013xba,Guzey:2013qza}; see Fig.~\ref{fig:GG}. Therefore, it is natural to assume that 
hadronic fluctuations
of the photon described by the distribution $P_h(\sigma)$ are similar to those for the nuclear gluon distribution. In this case, 
one can express the ratio of moments $\langle \sigma^k \rangle/\langle \sigma\rangle$ entering Eq.~(\ref{tre}) in terms of two effective cross sections $\sigma_2(x)$ and $\sigma_3(x)$~\cite{Frankfurt:2011cs},
\begin{equation}
\frac{\langle \sigma^2 \rangle}{\langle \sigma \rangle} = \sigma_2(x) \,, \quad 
\frac{\langle \sigma^3 \rangle}{\langle \sigma \rangle} = \sigma_3(x) \sigma_2(x) \,, 
\label{eq:sigma_eff}
\end{equation}
where we have explicitly indicated the dependence on Bjorken $x$. 
The $\sigma_2(x)$ cross section is unambiguously determined by the probability of diffraction in DIS 
on the proton in the gluon channel~\cite{H1:2006zyl} and the slope of the $t$ dependence of the $\gamma^{\ast} p \to X p$ cross section
($X$ denotes the diffractively produced final state), $B \simeq 6$ GeV$^{-2}$~\cite{H1:2006uea}.
At $x=10^{-3}$, one finds $\sigma_2(x)=25$ mb
with a relative error of approximately 15\% \cite{H1:2006uea,suppl} . 

On the other hand, the $\sigma_3(x)$ cross section needs to be modeled through $P_h(\sigma)$. Using two plausible models~\cite{Frankfurt:2011cs}, one 
finds $\sigma_3(x)=30-50$ mb. However, one of the key advantages of light nuclei compared to heavy nuclei is that
the sensitivity to the value of $\sigma_3$ is negligible
since the contribution of the
$k=3$ and $k=4$ terms in Eq.~(\ref{uno}) is very small in the studied range of $t$ (see the Supplemental Material~\ref{sec:suppl} 
for an analysis of this sensitivity).
In our calculations and in the plots, we used $\sigma_3(x)=35$ mb.
Thus, the effect of nuclear shadowing is determined by the interaction with two nucleons, whose strength is controlled by $\sigma_2(x)$.

At the same time, away from the minimum of the cross section and for large $|t|$, the three-body contribution becomes significant 
and reduces the cross section by approximately a factor of $2$. Hence, accurate measurements at large $|t|$ will allow one
to extract $\sigma_3(x)$ as well.

Finally, for the slope of the $t$ dependence the $\gamma^{\ast} p \to J/\psi p$ cross section, 
we use $B_0(x)=4.5$ GeV$^{-2}$,
with a relative error of approximately 10 \%, measured by H1 and ZEUS collaborations at HERA (see Ref.~\cite{Guzey:2013qza} for references).
This value corresponds to $x \simeq 10^{-3}$, typical for the EIC kinematics.
In addition, using the Gribov-Migdal relation, we estimate 
$\eta_0$ and $\eta$ 
by exploiting the measured energy dependence of the corresponding amplitudes:
$\eta_0=(\pi/2) \times 0.1 \simeq 0.16$ and $\eta=(\pi/2) \times 0.2 \simeq 0.3$.
In our analysis, we neglected the $t$ dependence of $\eta$ and $\eta_0$ since the slopes of the corresponding scattering
amplitudes weakly depend on energy [i.e., the slopes of the Regge trajectories $\alpha^{\prime}(0)$ are small].

The results are presented in Figs. 3-5,
taking into account the relative errors on $\sigma_2$ and $B_0$ discussed above. Notice that these uncertainties
do not affect our numerical predictions significantly.
In facts, the bulk of the predicted strong $t$ dependence is given by the nuclear $k$-body form factors, $\Phi_k$.
The latter quantities are
calculated with the most recent realistic potentials and the theoretical uncertainty on them, in the relevant
kinematical region, is very small. An example of the convergence of the nuclear calculation is provided in the Supplemental Material.

Figure~\ref{f_tre} shows our predictions for the ratio of the differential
cross section for $J/\psi$ coherent production on $^4$He to 
that for the nucleon 
at $t=0$
as a function of $-t$ at $x=10^{-3}$.
One can see from the figure that the cross section is dominated by the
one-body (IA) and the two-body rescattering
dynamics.
The first minimum is clearly shifted from
$-t=$ 0.45 GeV$^2$ to
$-t=$ 0.27 GeV$^2$, essentially due to the two-body contribution.
Since one-body dynamics is under remarkable theoretical control, 
it allows one to disentangle two-body dynamics
and unambiguously relate it to
leading-twist gluon nuclear shadowing.
Note also that the clear minimum of the $t$ dependence in the IA case is filled because $\eta_0 \neq \eta \neq 0$ 
in the full calculation. This represents a unique opportunity to measure 
the ratios of the real to imaginary parts of the corresponding scattering amplitudes.

The quality of the IA result can be tested at $x=0.05$,
where it is expected to 
be dominating in a broad range of $t$
due to a vanishingly small contribution of the shadowing correction. 
This is illustrated in
Fig.~\ref{fig_4}
presenting
the $x$ evolution of the gluon shadowing
correction in $^4$He.
It shows the ratio of the differential
cross section for $J/\psi$ coherent production on
$^4$He to the same quantity at $t=0$ as a function of
$-t$.
At $x=10^{-3}$, the full result is shown. At $x=0.05$,
  the IA result is presented. In the latter case, the parameters of the model
of $J/ \psi$ production have been properly changed (in particular, 
we used $B_0(x) = 3$ GeV$^{-2}$~\cite{Frankfurt:2010ea}). 

Note that this $x$ evolution of the $t$ dependence agrees with that predicted in Ref.~\cite{Frankfurt:2010ea},
 which was obtained considering HERA data; a check of this model will be possible at the EIC.
Since the 
one-body contribution dominates the cross section at $x=0.05$,
 where no shadowing is expected
in a wide range of $t$, the emergence of LT gluon shadowing
at lower $x$ points to a significant broadening in the impact parameter space of the nuclear gluon distribution, as discussed
in Ref.~\cite{Guzey:2016qwo}
for heavy nuclei. If confirmed, this observation
would be a relevant step towards a 3D imaging of gluons in nuclei.

We have also repeated our analysis  for the
$^3$He system, which will be systematically used at an EIC.
In this case, 
the nuclear ffs $\Phi_1, \Phi_2$, and $\Phi_3$
have been calculated using a realistic wave function developed along the lines
of Ref.~\cite{Marcucci:2019hml}
and using
the AV18 nucleon-nucleon potential~\cite{Wiringa:1994wb} and including
UIX three-body forces~\cite{Pudliner:1995wk}.
Again, as a consistency check of the numerics, the 
relation $\Phi_2(\vec q_2=0,\vec q_{1}) =
\Phi_1(\vec q_{1})$
has been successfully tested.

Figure~\ref{f_5} shows the ratio of the differential
cross section for $J/\psi$ coherent production on
$^3$He to the same quantity for the nucleon target 
at $t=0$ as a function of $-t$
at $x=10^{-3}$.
One can see from the figure that the pattern of the $t$ dependence is similar to that found for $^4$He.
Again, the one-body and two-body scatterings dominate the cross sections. A
relevant shift in the minimum momentum transfer due to two-body dynamics
is predicted and, since the IA is under theoretical control, there are very good chances to 
disentangle two-body dynamics, i.e., LT gluon shadowing,
from the one-body contribution.
In addition, accurate measurements at large $|t|$ would also allow one to extract the three-nucleon contribution.

One should also mention that, within our approach, the average number of participating nucleons 
is $\nu = A\sigma_{\gamma^{\ast} N}/ \sigma_{\gamma^{\ast} A}$~\cite{Bertocchi:1976bq}, which
leads to $\nu \sim 1.7$ for $x=10^{-3}$. Thus, using the set of $^3$He and $^4$He data, one would 
be able to 
test the consistency of our description.

\textbf{\textit{Conclusions}}.
Measurements of coherent $J/\psi$ electroproduction at finite values of the momentum
transfer $t$ with 
light ion beams at an EIC can nicely complement investigations
performed at the LHC with ultraperipheral collisions of heavy nuclei.
In particular, it will be possible to establish how many 
nucleons contribute to the impressive gluon shadowing
seen at the LHC, which constitutes important information hardly
accessible in the LHC data collected with heavy nuclei 
probing predominantly the $t \simeq 0$ values.
We demonstrated this by
performing a realistic calculation
for the $^3$He and $^4$He systems at $t \ne 0$ and considering
contributions coming from different numbers of nucleons involved
in the process. We have clearly shown that the first diffraction minimum is 
shifted with respect to that predicted by the IA calculation.
Since the latter contribution is  under good theoretical control, very good 
opportunities to disentangle multinucleon dynamics, in particular
two-nucleon dynamics contributing to gluon shadowing, are expected.
It should be possible to perform such measurements at the EIC, due to its projected high luminosity, designed for precision measurements of exclusive processes \cite{AbdulKhalek:2021gbh}.
An encouraging estimate of the events rate expected at the EIC
is presented in the Supplemental Material.
Besides, one should also note that the measurements planned at EIC for the free proton target, in particular, those of the slopes $B_0$ and $B$ (related to $\sigma_2$), will reduce the uncertainties of our results shown here.
It will also be possible to obtain unique information on the real part of the corresponding scattering amplitude.
Analyzing the $x$ evolution of the $t$ dependence predicted in our calculation, 
the emergence of LT gluon shadowing at low $x$
 points to a significant broadening of the gluon distribution in impact parameter space.
This is just an example of many possibilities offered by the
process under scrutiny towards a novel 3D imaging of gluons in nuclei.

The use of light ion beams would greatly expand the EIC potential for probing the small-$x$
dynamics. We will perform further investigations considering additional light ions
(e.g., deuteron beams), other vector mesons in the final state allowing for a sizable longitudinal momentum transfer,
deeply virtual Compton scattering, and the $Q^2$ dependence of cross sections of these processes.

{\textbf{\textit{Acknowledgements}}}.
The research of M.R. and S.S. is supported in part by the STRONG-2020 project of the European Union's Horizon 2020 research and innovation programme: Fund no 824093.
The research of M.S was supported by the US Department of Energy Office of Science, Office of Nuclear Physics under 
Award No.~DE-FG02-93ER40771.

\section{Supplemental material}
\label{sec:suppl}

In this document we present supplemental material for our paper 
to support the 
obtained results. In particular, an estimate of the uncertainties on our results, as well as
an estimate of the feasibility of measurements of $J/\psi$ photoproduction on light nuclei at EIC, will be detailed.

\textbf{\textit{Estimate of the accuracy of our calculations}}.
For the quantity
\begin{eqnarray}
R &=&
\frac{d \sigma_{\gamma^{\ast} A \rightarrow J/\psi A }}{dt}(t)
/
\frac{d \sigma_{\gamma^{\ast} N \rightarrow J/\psi N }}{d t}(t=0) \nonumber\\
&=& 
\left |F_1(t)e^{[B_{0}(x)/2]t}+
\sum_{k=2}^A F_k(t) \right |^2 \,,
\label{uno}
\end{eqnarray}
shown in Figs. 3-5,
the error propagation has been performed evaluating
\begin{eqnarray}
\Delta R = \sqrt{ 
\left ( \frac{\partial R}{\partial B_0}   \right )^2 (\Delta B_0)^2
+
\left ( \frac{\partial R}{\partial \sigma_2}   \right )^2 (\Delta \sigma_2)^2
}
\label{eq:err_prop}
\end{eqnarray}
with $\Delta B_0 = 0.1 \, B_0$ and $\Delta \sigma_2
= 0.15 \, \sigma_2$. Of course, the latter uncertainty does not affect the IA result. 

\begin{figure}[t]
\begin{center}
\includegraphics[scale=0.7]{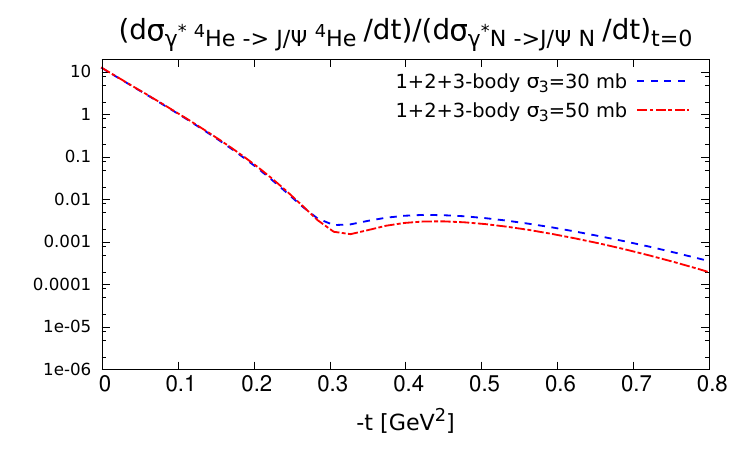}
\end{center}
\caption{Sensitivity to changes of the parameter $\sigma_3$ in its range of variation: the effect is seen only in the full result, 1b+2b+3b.}
\label{fig:sup1}
\end{figure}

The corresponding evaluated error bands can be seen in Figs. 3-5. 
In the figures, the value $\sigma_3=35$ mb  has been chosen.
The effect of the uncertainty on $\sigma_3$ is shown
in Fig.~\ref{fig:sup1}. 
Of course it enters only the 1b+2b+3b result. It is seen that such an error is negligible in the kinematical region of interest and therefore the uncertainty on $\sigma_3$ has not been included in the error propagation, Eq. (\ref{eq:err_prop}).

\begin{figure}[h]
\begin{center}
\includegraphics[scale=0.7]{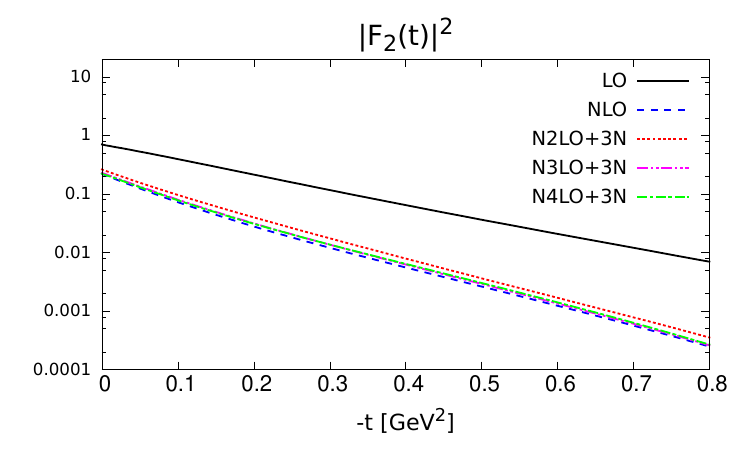}
\end{center}
\caption{Convergence of the nuclear calculation, from Leading Order to fifth Order in the chiral expansion. 
Three-body forces ("3N") arise naturally, starting at
N2LO.
The squared absolute value of the two-body form factor $F_2$ for $^4$He is shown as an example.}
\label{fig:sup2}
\end{figure}

To quantify the uncertainty associated with the nuclear dynamics, in Fig.~\ref{fig:sup2} we present as an example the function $|F_2(t)|^2$ for $^4$He 
(Eq.~(\ref{tre})), which is obtained from the two-body form-factor $\Phi_2$ (Eq.~(\ref{cinque}) and Fig.~\ref{f_due}) considering the nuclear wave functions obtained as solutions of the Schr\"odinger equation using effective interactions of growing accuracy, from Leading Order to the fifth order in the chiral expansion~\cite{Entem:2017gor}. 
It is seen that, in particular in the kinematical region of interest, the convergence is very fast and the use of even more accurate wave functions is not expected to produce an effect on our results and change appreciably our conclusions.
For this reason no uncertainty in 
$\Phi_k$ has been considered in Eq.~(\ref{eq:err_prop}) here above.

\textbf{\textit{Estimates of feasibility of measurements of $J/\psi$ photoproduction on light nuclei at EIC}}.
Using the standard equivalent photon approximation (EPA), the cross section of $J/\psi$ photoproduction in electron-nucleus scattering
$e+A \to e^{\prime}+\gamma+A \to e^{\prime}+J/\psi +A$ reads
\begin{equation}
\frac{d \sigma_{e+A \to e^{\prime}+J/\psi +A}}{dy dt} =N_{\gamma/e}(x) \frac{d\sigma_{\gamma+A \to J/\psi+A}}{dt}(t) \,,
\label{eq:cs1}
\end{equation}
where $N_{\gamma/e}(x)$ is the photon flux and $x$ is the momentum fraction of the electron carried by the photon;
$y$ and $t=-|\vec{p}_T|^2$ are the rapidity and transverse momentum squared of $J/\psi$, respectively.

For the photon flux, we use the standard expression, see, e.g.~\cite{Klasen:2002xb},
\begin{eqnarray}
N_{\gamma/e}(x) &=& \frac{\alpha_{e.m.}}{2 \pi} \Bigg[\frac{1+(1-x)^2}{x}\ln \frac{Q^2_{\rm max} (1-x)}{m_e^2 x^2} \nonumber\\
&+&2 m_e^2 x \left(\frac{1}{Q^2_{\rm max}}-\frac{1-x}{m_e^2 x^2} \right) \Bigg] \,,
\label{eq:flux}
\end{eqnarray}
where $m_e$ is the electron mass; $Q^2_{\rm max}=0.1$ GeV$^2$ is the maximal photon virtuality (the value is motivated 
by studies of photoproduction at HERA).

For the $d\sigma_{\gamma+A \to J/\psi+A}/dt$ photoproduction cross section, we used the results of our calculations.
In particular, we employed the following convenient relation
\begin{eqnarray}
\frac{d\sigma_{\gamma+A \to J/\psi+A}}{dt}(t) &=& \frac{d\sigma_{\gamma+p \to J/\psi+p}}{dt}(t=0) \nonumber\\
& \times &\frac{d\sigma_{\gamma+A \to J/\psi+A}/dt}{d\sigma_{\gamma+p \to J/\psi+p}/dt(t=0)}
 \,,
\label{eq:cs2}
\end{eqnarray} 
where for the cross section of elastic $J/\psi$ photoproduction on the proton
$d\sigma_{\gamma+p \to J/\psi+p}/dt(t=0)$, we used the H1 parametrization~\cite{Alexa:2013xxa};
the ratios $(d\sigma_{\gamma+A \to J/\psi+A}/dt)/(d\sigma_{\gamma+p \to J/\psi+p}/dt(t=0))$ are taken from Fig.~\ref{f_tre} 
(He-4) and Fig.~\ref{f_5} (He-3).

In the lab frame, the photon energy $\omega$ is related to the $J/\psi$ rapidity as follows (the rapidity is defined along the 
electron beam direction)
\begin{equation}
\omega=\frac{M_{J/\psi}}{2} e^{y} \,,
\label{eq:omega}
\end{equation}
where $M_{J/\psi}$ is the mass of $J/\psi$. Therefore, the momentum fraction $x$ is $x=\omega/E_{e}$, where $E_e$ is the electron beam energy.

In our estimate, we use the following inputs :
\begin{itemize}
\item
$E_e=20$ GeV is the electron beam energy,
typical at EIC \cite{AbdulKhalek:2021gbh} 
\item
$E_A=100$ GeV is the nuclear beam energy (per nucleon) \cite{AbdulKhalek:2021gbh}.
\item
${\cal L}_{\rm int}=10$ fb$^{-1}$ is the integrated luminosity~\cite{AbdulKhalek:2021gbh}.
\item
$Br(J/\psi \to \mu^{+} \mu^{-})=5.93$\% is the $J/\psi \to \mu^{+} \mu^{-}$ branching ratio.
\end{itemize}

Therefore, the number of events in a given bin of $\Delta y$ and $\Delta t$ is 
\begin{eqnarray}
\Delta N &=&{\cal L}_{\rm int} \frac{d \sigma(e+A \to e^{\prime}+J/\psi +A)}{dy dt} \Delta y \Delta t \nonumber\\
& \times& Br(J/\psi \to \mu^{+} \mu^{-})
\,.
\label{eq:N}
\end{eqnarray}
We assume that $\Delta y=1$ and $\Delta t=0.1$ GeV$^2$.
We find that $\Delta N$ and the associated statistical error $1/\sqrt{\Delta N}$ depend on $y$ and $t$.
In the region of positive $y$ corresponding to large $\omega$ and small Bjorken $x \sim 10^{-3}$, $1/\sqrt{\Delta N} \leq 5$\% 
in the entire studied region of $t$, $0 \leq |t| < 0.8$ GeV$^2$. 
At central and backward rapidities corresponding to $x_B \sim 0.05$, $1/\sqrt{\Delta N} \leq 2$\% and better in the studied 
region $0 \leq |t| < 0.8$ GeV$^2$.


\begin{thebibliography}{99}

\bibitem{Frankfurt:1988nt}
L.~L.~Frankfurt and M.~I.~Strikman,
Phys. Rept. \textbf{160}, 235-427 (1988).

\bibitem{Arneodo:1992wf}
M.~Arneodo,
Phys. Rept. \textbf{240}, 301-393 (1994).

\bibitem{Piller:1999wx}
G.~Piller and W.~Weise,
Phys. Rept. \textbf{330}, 1-94 (2000)
[arXiv:hep-ph/9908230 [hep-ph]].

\bibitem{Armesto:2006ph}
N.~Armesto,
J. Phys. G \textbf{32}, R367-R394 (2006)
[arXiv:hep-ph/0604108 [hep-ph]].

\bibitem{Frankfurt:2011cs}
L.~Frankfurt, V.~Guzey and M.~Strikman,
Phys. Rept. \textbf{512}, 255-393 (2012)
[arXiv:1106.2091 [hep-ph]].

\bibitem{Glauber:1955qq}
R.~J.~Glauber,
Phys. Rev. \textbf{100}, 242-248 (1955).

\bibitem{Glauber:1970jm}
R.~J.~Glauber and G.~Matthiae,
Nucl. Phys. B \textbf{21}, 135-157 (1970).

\bibitem{Mandelstam:1963cw}
S.~Mandelstam,
Nuovo Cim. \textbf{30}, 1148-1162 (1963).

\bibitem{Gribov:1968jf}
V.~N.~Gribov,
Sov. Phys. JETP \textbf{29} (1969), 483-487
(Zh. Eksp. Teor.Fiz. 56 (1969) 892-901).

\bibitem{Salgado:2011wc}
C.~A.~Salgado, J.~Alvarez-Muniz, F.~Arleo, N.~Armesto, M.~Botje, M.~Cacciari, J.~Campbell, C.~Carli, B.~Cole and D.~D'Enterria, \textit{et al.}
J. Phys. G \textbf{39}, 015010 (2012)
[arXiv:1105.3919 [hep-ph]].

\bibitem{Citron:2018lsq}
Z.~Citron, A.~Dainese, J.~F.~Grosse-Oetringhaus, J.~M.~Jowett, Y.~J.~Lee, U.~A.~Wiedemann, M.~Winn, A.~Andronic, F.~Bellini and E.~Bruna, \textit{et al.}
CERN Yellow Rep. Monogr. \textbf{7}, 1159-1410 (2019)
[arXiv:1812.06772 [hep-ph]].

\bibitem{Accardi:2012qut}
A.~Accardi, J.~L.~Albacete, M.~Anselmino, N.~Armesto, E.~C.~Aschenauer, A.~Bacchetta, D.~Boer, W.~K.~Brooks, T.~Burton and N.~B.~Chang, \textit{et al.}
Eur. Phys. J. A \textbf{52}, no.9, 268 (2016)
[arXiv:1212.1701 [nucl-ex]].

\bibitem{Frankfurt:1998ym}
L.~Frankfurt and M.~Strikman,
Eur. Phys. J. A \textbf{5}, 293-306 (1999)
[arXiv:hep-ph/9812322 [hep-ph]].

\bibitem{Guzey:2013xba} 
  V.~Guzey, E.~Kryshen, M.~Strikman and M.~Zhalov,
  Phys.\ Lett.\ B {\bf 726}, 290 (2013)
  [arXiv:1305.1724 [hep-ph]].
  
\bibitem{Guzey:2013qza} 
  V.~Guzey and M.~Zhalov,
  JHEP {\bf 1310}, 207 (2013)
  [arXiv:1307.4526 [hep-ph]].
  
\bibitem{Guzey:2020ntc}
V.~Guzey, E.~Kryshen, M.~Strikman and M.~Zhalov,
Phys. Lett. B \textbf{816} (2021), 136202
[arXiv:2008.10891 [hep-ph]].

\bibitem{Lappi:2013am}
T.~Lappi and H.~Mantysaari,
Phys. Rev. C \textbf{87}, no.3, 032201 (2013)
[arXiv:1301.4095 [hep-ph]].

\bibitem{Ryskin:1992ui}
M.~G.~Ryskin,
Z. Phys. C \textbf{57}, 89-92 (1993)

\bibitem{Brodsky:1994kf}
S.~J.~Brodsky, L.~Frankfurt, J.~F.~Gunion, A.~H.~Mueller and M.~Strikman,
Phys. Rev. D \textbf{50}, 3134-3144 (1994)
[arXiv:hep-ph/9402283 [hep-ph]].

\bibitem{Guzey:2016qwo}
V.~Guzey, M.~Strikman and M.~Zhalov,
Phys. Rev. C \textbf{95} (2017) no.2, 025204.
[arXiv:1611.05471 [hep-ph]].


\bibitem{Levin:1975pc}
E.~M.~Levin and M.~I.~Strikman,
Sov. J. Nucl. Phys. \textbf{23} (1976), 216
LENINGRAD-75-203.

\bibitem{Frankfurt:2003jf}
L.~Frankfurt, V.~Guzey and M.~Strikman,
Phys. Rev. Lett. \textbf{91}, 202001 (2003)
[arXiv:hep-ph/0304149 [hep-ph]].


\bibitem{Good:1960ba}
M.~L.~Good and W.~D.~Walker,
Phys. Rev. \textbf{120}, 1857-1860 (1960).

\bibitem{Blaettel:1993ah}
B.~Blaettel, G.~Baym, L.~L.~Frankfurt, H.~Heiselberg and M.~Strikman,
Phys. Rev. D \textbf{47}, 2761-2772 (1993).


\bibitem{Bujak:1978rla}
A.~Bujak, A.~Kuznetsov, B.~Morozov, V.~A.~Nikitin, P.~Nomokonov, Y.~Pilipenko, V.~Smirnov, E.~Jenkins, E.~Malamud and M.~Miyajima, \textit{et al.}
Phys. Rev. D \textbf{23} (1981), 1895.

\bibitem{AbdulKhalek:2021gbh}
R.~Abdul Khalek, A.~Accardi, J.~Adam, D.~Adamiak, W.~Akers, M.~Albaladejo, A.~Al-bataineh, M.~G.~Alexeev, F.~Ameli and P.~Antonioli, \textit{et al.}
[arXiv:2103.05419 [physics.ins-det]].

\bibitem{Marcucci:2019hml}
L.~E.~Marcucci, J.~Dohet-Eraly, L.~Girlanda, A.~Gnech, A.~Kievsky and M.~Viviani,
Front. in Phys. \textbf{8} (2020), 69
[arXiv:1912.09751 [nucl-th]].

\bibitem{Entem:2017gor}
D.~R.~Entem, R.~Machleidt and Y.~Nosyk,
Phys. Rev. C \textbf{96}, no.2, 024004 (2017)
[arXiv:1703.05454 [nucl-th]].

\bibitem{H1:2006zyl}
A.~Aktas \textit{et al.} [H1],
Eur. Phys. J. C \textbf{48}, 715-748 (2006)
[arXiv:hep-ex/0606004 [hep-ex]].

\bibitem{H1:2006uea}
A.~Aktas \textit{et al.} [H1],
Eur. Phys. J. C \textbf{48}, 749-766 (2006)
[arXiv:hep-ex/0606003 [hep-ex]].

\bibitem{suppl} See Supplemental Material at http://link.aps.org/supplemental/10.1103/PhysRevLett.129.242503, which includes
Refs. \cite{Klasen:2002xb,Alexa:2013xxa}, for an estimate of the accuracy of our calculation and on the feasibility of the measurements under scrutiny at the EIC.
\bibitem{Klasen:2002xb}
M.~Klasen,
Rev. Mod. Phys. \textbf{74}, 1221-1282 (2002)
[arXiv:hep-ph/0206169 [hep-ph]].

\bibitem{Alexa:2013xxa}
C.~Alexa \textit{et al.} [H1],
Eur. Phys. J. C \textbf{73} (2013) no.6, 2466
[arXiv:1304.5162 [hep-ex]].

\bibitem{Frankfurt:2010ea}
L.~Frankfurt, M.~Strikman and C.~Weiss,
Phys. Rev. D \textbf{83}, 054012 (2011)
[arXiv:1009.2559 [hep-ph]].


\bibitem{Wiringa:1994wb}
R.~B.~Wiringa, V.~G.~J.~Stoks and R.~Schiavilla,
Phys. Rev. C \textbf{51}, 38-51 (1995)
[arXiv:nucl-th/9408016 [nucl-th]].

\bibitem{Pudliner:1995wk}
B.~S.~Pudliner, V.~R.~Pandharipande, J.~Carlson and R.~B.~Wiringa,
Phys. Rev. Lett. \textbf{74}, 4396-4399 (1995)
[arXiv:nucl-th/9502031 [nucl-th]].

\bibitem{Bertocchi:1976bq}
L.~Bertocchi and D.~Treleani,
J. Phys. G \textbf{3}, 147 (1977)

\end{thebibliography}
\end{document}